\documentclass[aps,prb,twocolumn,showpacs,letterpaper,superscriptaddress]{revtex4-1}
\usepackage{amsmath}
\usepackage{comment}
\usepackage{color}
\usepackage[usenames,dvipsnames]{xcolor}
\usepackage{soul} 
\usepackage{graphics}\usepackage{epsfig}\usepackage{subfigure}
\usepackage{hyperref} 
\usepackage{multirow}
\usepackage{amsmath}
\usepackage{tabularx}

\begin{document}

\title{Doping the spin-polarized Graphene minicone on Ni(111)}

\author{Cesare Tresca}
\email[]{cesare.tresca@spin.cnr.it}
\affiliation{CNR-SPIN c/o Dipartimento di Scienze Fisiche e Chimiche, Università degli Studi dell’Aquila, Via Vetoio 10, I-67100 L’Aquila, Italy}
\author{Gianni Profeta}
\affiliation{CNR-SPIN c/o Dipartimento di Scienze Fisiche e Chimiche, Università degli Studi dell’Aquila, Via Vetoio 10, I-67100 L’Aquila, Italy} 
\affiliation{Dipartimento di Scienze Fisiche e Chimiche, Università degli Studi dell’Aquila, Via Vetoio 10, I-67100 L'Aquila, Italy} 
\author{Federico Bisti}
\email[]{federico.bisti@univaq.it}
\affiliation{Dipartimento di Scienze Fisiche e Chimiche, Università degli Studi dell’Aquila, Via Vetoio 10, I-67100 L'Aquila, Italy}

\begin{abstract}
In the attempt to induce spin-polarized states in graphene, rare-earth deposition on Gr/Co(0001) has been demonstrated to be a successful strategy: the coupling of graphene with the cobalt substrate provides spin-polarized conical-shaped states (mini-cone) and the rare-earth deposition brings these states at the Fermi level.
In this manuscript we theoretically explore the feasibility of an analogue approach applied on Gr/Ni(111) doped with rare-earth ions. Even if not well mentioned in the lecture also this system owns a mini-cone, similar to the cobalt case. By testing different rare-earth ions, not only we suggest which one can provide the required doping but we explain the effect behind this proper charge transfer.
\end{abstract}



\maketitle
\section{Introduction}

\begin{table*}[t]
\caption{Structural and magnetic properties of the studied systems: the average distance between graphene and the Ni(111) surface (d$_{C-Ni}$); the rippling of the graphene flake ($\Delta$d$_{C}$); the average distance between the RE adatom and the carbon atoms (d$_{RE}$); the predicted atomic magnetic moments for the top-most Ni atoms (m$_{Ni^*}$), for the carbon atoms (m$_C$ for A and B sites respectively) and for the RE atoms (m$_{RE}$). 
\label{tab1}}
\begin{tabularx}{0.75\textwidth}{cccccc}
\hline
 & Gr/Ni(111)   & La & Eu & Gd & Yb\\
\hline
d$_{C-Ni}$ (\AA)   &  2.05  & 2.20   & 2.12  & 2.21   &  2.12  \\ 
$\Delta$d$_{C}$ (\AA)   &  0.01  & 0.05    &  0.04 & 0.05   &  0.05  \\
d$_{RE}$   (\AA) & --   &  2.32   & 2.38  & 2.18  &  2.23  \\ 
m$_{Ni^*}$ ($\mu_B$)   & 0.52 &  0.54   &  0.35   & 0.55   & 0.34  \\ 
m$_C$ ($\mu_B$)   & 0.03 / -0.01(5) &   0.00  & 0.00 / -0.00(7)    &   -0.00(6) / 0.00 &  0.00 / -0.00(8) \\ 
m$_{RE}$ ($\mu_B$)  & --     &  0.16    & -7.05 & 7.37 &  -0.00(8)  \\ 
\hline
\end{tabularx}
\end{table*}

\begin{figure*}[t]
 \includegraphics[width=0.75\textwidth]{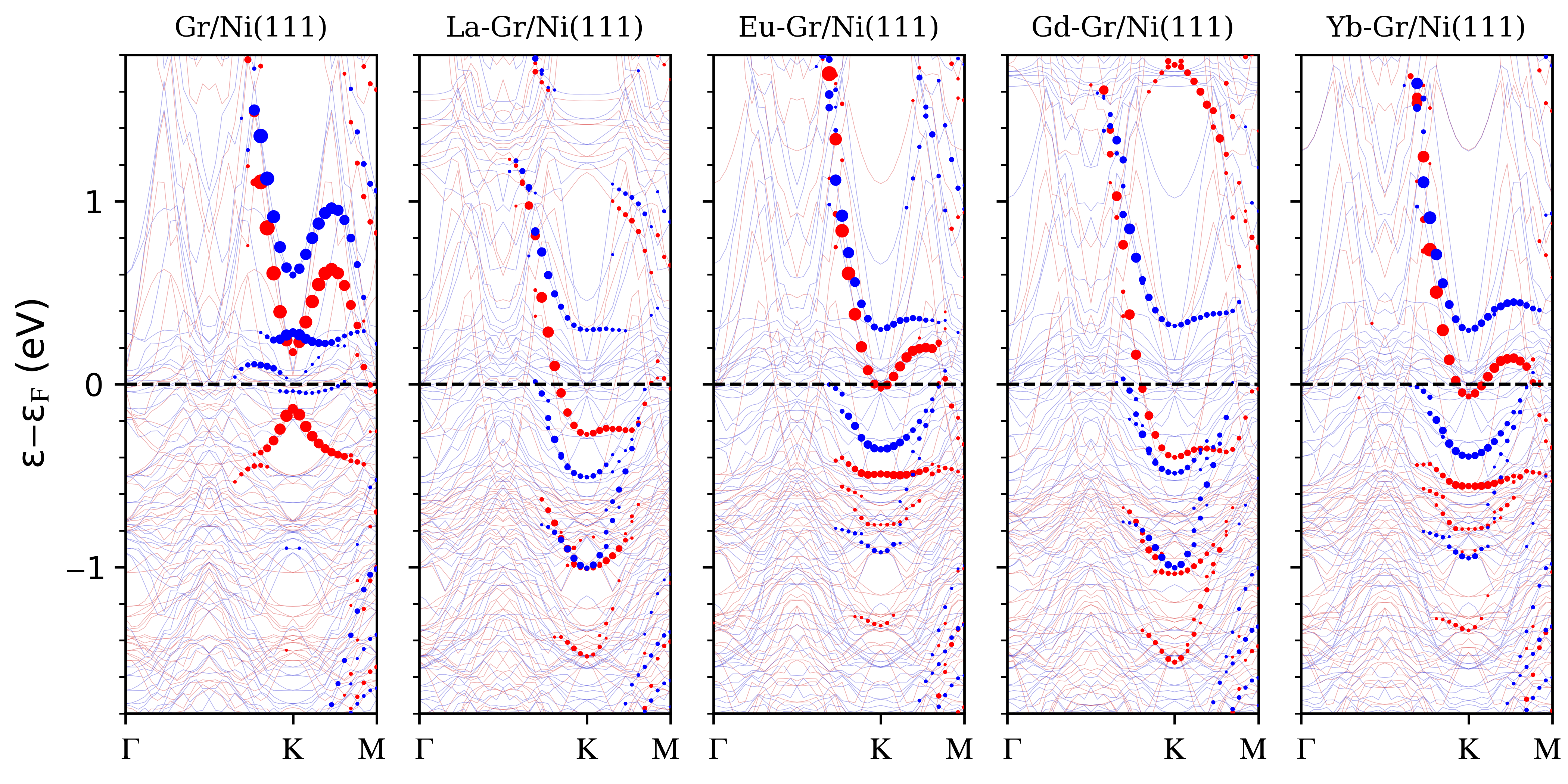}
 \caption {From left to right, with blue (red) lines the majority (minority) spin channel band structure of the pristine system followed by La, Eu, Gd and Yb adatoms on Gr/Ni(111). The band dispersions are along the $\Gamma-K-M$ direction of the graphene 1$\times$1 BZ. The size of blue (red) circles is proportional to the projection of the respectively eigen-function, from the majority (minority) spin channel, on the carbon $p_z$-states further unfolded on the graphene 1$\times$1 periodicity.\label{bands}
 }
\end{figure*}

The exceptional properties of graphene\cite{Novoselov2004,Novoselov2005} has stimulated studies for engineering the Dirac bands, giving rise to the stabilization of new and interesting electronic phases.
Exceptional examples are the superconducting phase\cite{Profeta2012,Ludbrook2015}, magnetic systems\cite{Ahn2020,PhysRevX.12.021010} for spintronic applications, heterostructures showing new properties, twisted and  Moir\'{e} configurations\cite{Cao2018}, nanostructures\cite{Senkovskiy2018,Liu2023} and photonic applications in linear and non-linear optics\cite{Kumar2021, PhysRevA.90.033828}, opening new routes for technological applications\cite{Schwierz2010,Ohno2010,Yang2017,Gargiani2017,Zhao2024}.
At the moment the most promising  ways to tailor the graphene physical properties is by means of hetero-atoms deposition, substrate induced interaction thought electronic and structural modifications, twisted and/or heterostructures formed with 2D systems\cite{Geim2013}.

In particular, its growth or transfer onto different substrates has been the subject of massive investigations in the past, with the aim of understanding how its electronic properties change from the isolated picture ('free-standing') as a result of interaction with the substrate\cite{PhysRevB.79.195425,PhysRevB.64.035405,Varykhalov2008,PhysRevB.80.035437,PhysRevB.79.045407,PhysRevX.2.041017,PhysRevB.88.235430,Tresca_2017,Gargiani2017} or following atomic intercalation and doping\cite{Fedorov2014,Verbitskiy2016,Bisti2021}.

The most important parameters defining the final graphene electronic properties are the lattice matching and the degree of hybridization of $\pi$ bands of graphene with the substrate.
For examples, the graphene grown on substrates with large lattice mismatch as Ir(0001) or Ru(0001) retains the linear $\pi$-bands dispersion close to the Fermi level without appreciable doping.
On the contrary, lattice matched Ni(111) and Co(0001) substrates strongly interact with graphene electrons while exhibiting spin-polarized bands\cite{Usachov2015,Tresca_2017, Jugovac2023, Weser2011,Gargiani2017}. In such strongly interacting scenario the peculiar high electronic charge mobility coming from the almost linear dispersion of the Dirac bands could be compromised if them are completely destroyed by the interaction with the substrate. 

This is not the case for Graphene growth on Co(0001). Indeed, even if a gap at the Dirac point of $\sim$0.4~eV is gained from the interaction with the substrate, the carbon $\pi$-bands are still highly dispersing bands (commonly called "minicone")\cite{Jugovac2023eu, Usachov2015}.

These features can bring to relevant spintronics applications once combined to the already known capability of graphene to sustain spin currents once injected by spin-polarized electrodes\cite{Tombros2007}, if low spin-orbit coupling, negligible hyperfine interaction, and gate tunability are preserved.
A fundamental element to consider is that the minicone in graphene/Co(0001) results to have the lower part of splitted cones fully occupied and the upper part fully unoccupied, determining a null contribution in the electric conduction. Finding a process that may partially occupy these minicones is therefore necessary in order to take advantage of them. One such mechanism is the deposition of dopant adatoms, which increases the electron charge on the carbon layer.
Although this procedure has been efficiently done for quasi-free-standing graphene\cite{Khademi2016,Link2019,Tresca2018,Verbitskiy2016, Fedorov2014,Rosenzweig2020,Ehlen2020, Bisti2021,Jugovac2023eu,PhysRevLett.132.266401,PhysRevResearch.5.013099}, in the present case it is fundamental to avoid intercalation of the dopants adatoms between substrate and the graphene adlayer. In fact, intercalation tend to detach graphene from the substrate\cite{Bisti2015,PhysRevB.105.L241107,Jugovac2023,PhysRevLett.132.266401}, destroying the spin-polarized states induced by the strong hybridization with the magnetic substrate.

Very recently, low temperature deposition of dopants, namely Europium adatoms on graphene/Co(0001) was demonstrated to be an effective technique to adsorb the dopants on graphene sheet in an ordered $\sqrt{3}\times\sqrt{3}$-R$30^\circ$  reconstruction, without intercalation and thus heavily doping the minicone keeping its peculiar spin-polarization\cite{Jugovac2023eu}.
This finding paves the way for the search of other magnetic substrates where to observe an analogue mechanism.

A good candidate is Ni(111): the electronic band dispersion of graphene growth on Ni(111) shows the presence of the spin-polarized minicone\cite{Voloshina2011,Guo2022,Tresca_2017} in analogy with graphene on Co(0001), guaranteeing a magnetic ordering above room temperature\cite{PhysRevLett.132.266401}.

At the same time, if the intercalation is precluded (as by deposition at low temperature), RE adatoms are expected to reconstruct into an ordered surface on top of graphene transferring electronic charge to it, as demonstrated in the case of Eu on graphene/Co(0001)\cite{Jugovac2023eu}.

The scope of the present manuscript is precisely to examine the doping mechanism induced by rare-earth (RE) deposition on the graphene/Ni(111) minicone using first-principles density functional theory (DFT) calculations.
The considered RE (RE=La, Eu, Gd, Yb) adatoms are expected in a +2 configuration, and adsorbed on graphene in a $\sqrt{3}\times\sqrt{3}$-R30$^\circ$. 
The chosen RE adatoms provide a comprehensive picture on the influence of the different adatoms electronic configuration on the minicone doping and dispersion highlighting the relevant role of RE-$d$-states.

Finally, the calculations are expanded outside the domain of rare earth to  illustrate that analogues mechanisms can also be traced back in Lu and Y,  
demonstrating the wider valence of the presented concepts.

\begin{figure*}[t]
 \includegraphics[width=0.8\linewidth]{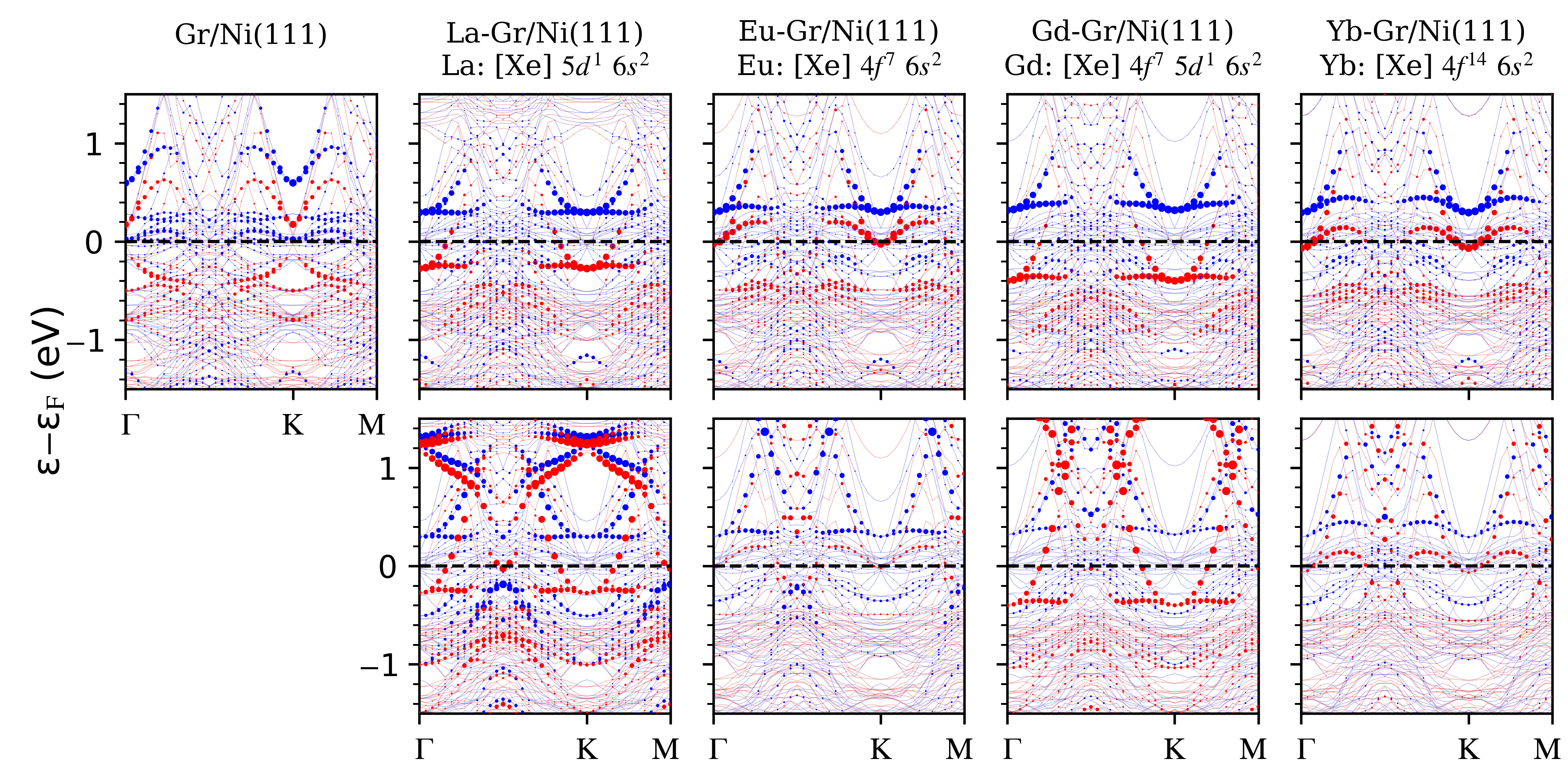}
 \caption {First row: electronic band structures for the studied systems projected on the surface Ni(d$_{z^2-r^2}$) states (top) and on the RE(d$_{xy}+$d$_{x^2-y^2}$) ones (bottom), from left to right, we show the pristine system followed by La, Eu, Gd and Yb adatoms on Gr/Ni(111). Blue and red are the majority and minority spin channels respectively. The RE atomic electronic configuration is reported.\label{bands_NiRe}
 }
\end{figure*}

\section{Computational methods}\label{Comp_det}
Theoretical calculations were performed using the Vienna ab-initio simulation package (VASP)\cite{Kresse1996}, using the generalized gradient approximation in the revised Perdew-Burke-Ernzerhof version (PBEsol)\cite{Perdew2008} for the exchange-correlation energy.
We used projected augmented-wave pseudopotentials\cite{Blochl1994}  for all the atomic species involved, with an energy cutoff up to 500~eV. 
The surfaces were simulated within a supercell approach which considers 6 Ni layers along the [111] direction and about $20$~\AA\ of vacuum. 

Graphene was adsorbed on the topmost Ni surface layer at the  experimental lattice parameter $2.49$~\AA, in the $1\times1$ reconstruction with the top-fcc stacking.
The ferromagnetic (FM) solution was considered for the Ni atoms in the calculations, while different spin-configurations were considered for the  magnetic adatoms adsorbed on the hollow site of graphene\cite{PhysRevB.82.245408} in a 
$\sqrt{3}\times\sqrt{3}$-R30$^\circ$ reconstruction.
Integration of charge density over the two dimensional Brillouin zone (BZ) was performed using an uniform 6$\times$6 Monkhorst and Pack grid\cite{Monkhorst1976} with a Gaussian smearing parameter $\sigma=0.05$~eV.

Total energy minimization was performed for all the atoms except for the bottom most four Ni layers that were fixed to their Ni bulk positions. 
The DFT+U approximation was adopted for the treatment of the $f$-orbitals, with the U and J parameters chosen in agreement with literature (for Eu and Gd we adopted U=5.9~eV, J=0.9~eV\cite{Jugovac2023}; for Yb U=2.0~eV, J=0.7~eV\cite{Guo2022}, while for La, Lu and Y no correction is needed).

\begin{figure*}[t]
 \includegraphics[width=0.7\linewidth]{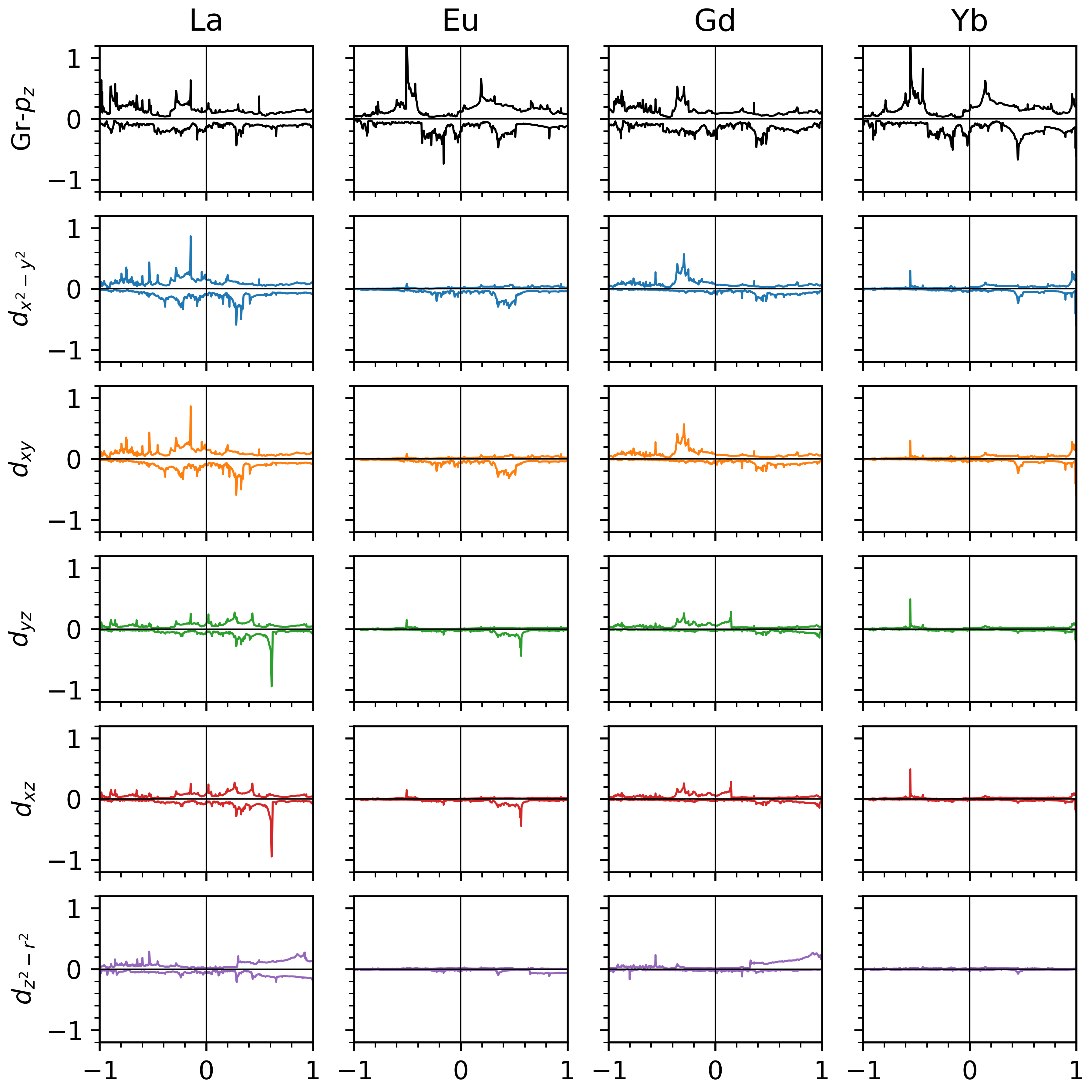}
 \caption {Selected orbital projected density of states (DOS) for the studied systems. From top to bottom we report the graphene p$_z$-projected density of states followed by the RE-d-projected DOS. In the abscissa we report the energies with respect to the $\varepsilon_F$ in eV, in the ordinate is shown the DOS expressed in states/eV/spin: majority spin in the positive axis, minority spin in the negative one.
 \label{figdos}
 }
\end{figure*}

\section{Results}

We start our study with graphene/Ni(111) system, taken as a reference. The adsorption distance between graphene and the Ni surface is predicted to be 2.05\AA\, in agreement with previous studies\cite{Tresca_2017,Guo2022}. Nickel substrate induces a small magnetic moments on the carbon atoms, showing an antiferromagnetic order between nonequivalent carbon sites (carbon in the on-top position has a magnetic moment aligned with those of the  Ni substrate (see Tab.\ref{tab1}).

Once the RE-adatom is included in the calculation, its adsorption causes the increase of the  graphene vertical distance  from the Ni(111) substrate with respect to the undoped case, regardless of the RE-atom involved (see Tab.\ref{tab1}). This effect is the natural consequence of the electronic charge transfer from the RE-atom to graphene, as it will be shown in the electronic band structure (see below). The adsorption distance between RE-atoms and the carbon layer is larger (of about 2.2\AA) for La and Gd than in the case of Eu and Yb (of about 2.1\AA).

\begin{figure*}[t]
 \includegraphics[width=0.75\textwidth]{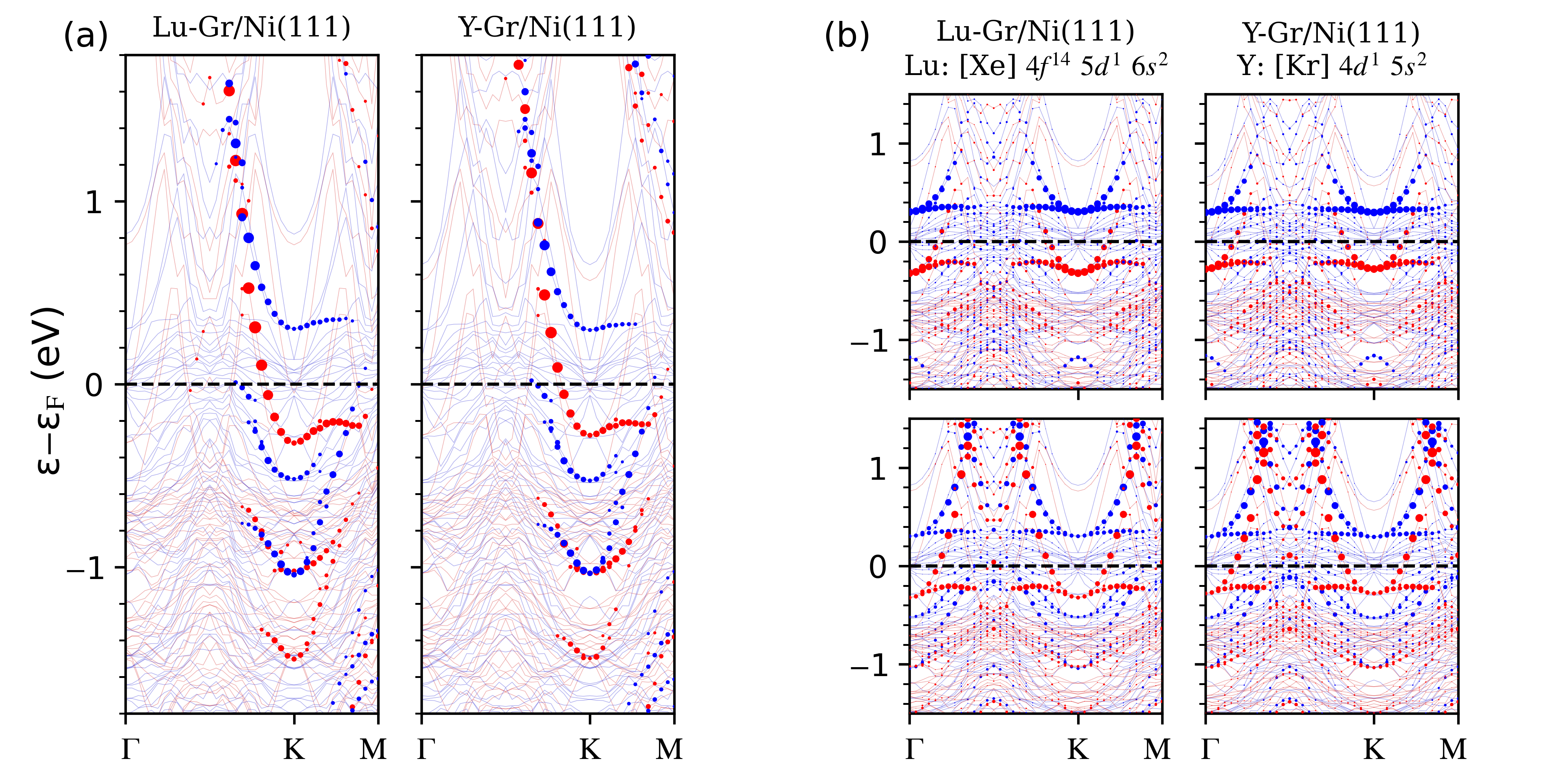}
\caption{Panel (a): unfolded electronic band structures projected on the carbon states (the size of circles is proportional to the spectral weight), for Lu (left) and Y (right) adatoms on Gr/Ni(111).
Panel (b): band structures for the considered systems projected on the surface
Ni($d_{z^2-r^2}$) states (top) and on the adatom ($d_{xy}+d_{x^2-y^2}$) ones (bottom), for Lu (left) and Y (right) cases.
Blue and red are the majority and minority spin channels, respectively. The adatom electronic configuration is reported.}
\label{bands2}
\end{figure*}

The structural properties correlate with the magnetic interaction between RE and the substrate: to the large (small) graphene-substrate distance of La and Gd (Eu and Yb) corresponds a ferromagnetic (anti-ferromagnetic) arrangement of them with respect to the Ni substrate (even if Yb shown a very poor residual magnetization). Such ordering of the adatoms influences the magnetic moment of the last Ni layer at the surface, giving an enhancement (reduction) in the ferromagnetic (anti-ferromagnetic) configuration (see Tab.\ref{tab1}).
In all the considered cases, the fragile magnetic ordering present in carbon atoms in Gr/Ni(111) system is practically destroyed by the presence of RE adatoms, due to the increased graphene-substrate distance.

The analysis of the electronic properties reveals the origin of these different behaviours. In Fig.\ref{bands} we report the spin polarized band structures for the considered systems, unfolded on the graphene BZ 1$\times$1 cell and projected on the C-$p_z$ orbitals to facilitate the recognition of the most dispersive graphene bands.

In line with previous studies\cite{Tresca_2017,Guo2022}, both majority and minority spin components exhibit a gap opening at the Dirac point, resulting in the so-called "minicone" shape at the K-point. Only the majority spin-component has occupied valence states, separated by an energy gap of $\sim0.35$~eV  from the conduction valley (see red dots in Fig.\ref{bands}).
The minicone gap opening originates from two main effects: the sublattice asymmetry induced by the Ni(111) substrate and the exchange field due to the strong $p-d$ hybridization coming from the spin-splitted Ni $d$-orbitals.

The adsorption of RE atoms drastically changes the electronic properties of the system: it induces electron doping, shifting downwards the spin-polarized carbon bands, and it brings strong modification of the graphene minicone producing an overall flattening of the band (in particular along the K-M direction, see below). 

Similarities in La and Gd behaviour opposed to the Eu and Yb cases are recognizable. First of all, the former provide an higher doping regime (downward shift of -0.6 eV) than the latter (-0.2 eV). Then, such higher doping present in La and Gd is accompanied by a strong flattening of the majority spin band dispersion along the K-M direction placing it below the Fermi level at about -0.25~eV. Whether instead in the case of Eu and Yb the graphene conduction majority spin band is almost unaltered in shape. All these effects can be connected to the valence configuration of the RE atom involved. 

In fact, both La and Gd have a $d$ states in valence (electronic configurations are $[Xe] \: 5d^1 \: 6s^2$ and $[Xe] \: 4f^7 \: 5d^1 \: 6s^2$ respectively), and those states tend to bond with graphene $p_z$.
An hybridization of the RE-$d_{xy/x^2-y^2}$ orbitals with the C-$p_z$ ones can completely disrupt the minicone structure, giving rise to a semi parabolic dispersion along the $\Gamma K$ path. In addition the band becomes extremely flat from K to M and, via a super-exchange mechanism, the RE results in a magnetic moment aligned parallel to the Ni(111) substrate.

To better clarify this effect, in Fig.\ref{bands_NiRe} we report the surface-Ni and RE $d$-projected states. From the first row in  Fig.\ref{bands_NiRe} it is evident as the hybidrization between the C-$p_z$ and Ni-$d_{z^2-r^2}$ orbitals underlies the spin-polarized state at the Fermi level both in the pristine system and in the case of RE adsorption cases.

At the same time, from the second row in Fig.\ref{bands_NiRe}, we note the presence of the RE-5$d_{xy,x^2-y^2}$ states hybridised with the C-$p_z$ (and Ni-$d_{z^2-r^2}$) ones only for La and Gd. Therefore, the spin-polarized electronic state at the Fermi level results extremely extended in real space: from the RE up to the surface-Ni layer in the out-of-plane direction.

A counter-proof of the hybridisation between the RE-$d_{xy,x^2-y^2}$ states with the and C-$p_z$ orbitals comes from the analysis of the projected densities of states reported in Fig.\ref{figdos}. As shown, only for La and Gd adsorption we have a perfect superposition of the C-$p_z$ and RE-$d_{xy,x^2-y^2}$ states for the "up"-spin channel. A negligible contribution from the other $d$-states is present.

To further expand the investigation of this effect, we thus consider the last of Lanthanides (Lutetium) and the first of transition metals (Yttrium), having respectively a filled $f$ orbital with a 5$d^1$ 6$s^2$ valence configuration (Lu) and a similar 4$d^1$ 5$s^2$ environment for Y (without $f$-states).

In agreement with the already observed behaviour, also Lu and Y are capable to "detach" graphene from the Ni(111) surface moving to a distance of 2.20\AA. A residual magnetization on the adatoms of 0.05 and 0.10$\mu_B$ respectively is present, ferromagnetically aligned with the Ni(111) substrate. The last Ni surface atoms exhibit a magnetization of 0.54~$\mu_B$, in both Lu and Y cases, in complete analogy with the previously considered RE-5$d$ cases (La and Gd). 
The structural details for these last system are summarized in in Tab.\ref{tab2}.

\begin{table}[h]
\caption{Structural and magnetic properties of the studied systems. We report the average distance between graphene and the Ni(111) surface (d$_{C-Ni}$), the rippling of the graphene flake ($\Delta$d$_{C}$) and the average distance between the RE adatom and the carbon atoms (d$_{RE}$). We also report the predicted atomic magnetic moments for the top-most Ni atoms (m$_{Ni^*}$), for the carbon atoms (m$_C$ for A and B sites respectively) and for the RE atoms (m$_{RE}$).\label{tab2} 
}
\begin{tabularx}{0.9\linewidth}{ccc}
\hline
 & Lu-Gr/Ni(111) &  Y-Gr/Ni(111)\\
\hline
d$_{C-Ni}$ (\AA)   & 2.20    &  2.20   \\ 
$\Delta$d$_{C}$ (\AA)    &  0.07   &  0.06   \\
d$_{RE}$   (\AA) & 2.08   &  2.14  \\ 
m$_{Ni^*}$ ($\mu_B$)      & 0.54   & 0.54  \\ 
m$_C$ ($\mu_B$)   &   -0.00(2) / 0.00(7) &  0.00 / 0.00(5) \\ 
m$_{RE}$ ($\mu_B$) & 0.05 &  0.10  \\ 
\hline
\end{tabularx}
\end{table}

From Fig.\ref{bands2} the electronic properties of the systems are essentially indistinguishable between the Lu or Y adsorption cases. The adatom $d$-states interact with C-$p_z$ ones, destroying the minicone, in perfect agreement with what happens in the already considered cases of La and Gd adsorption. Thus we conclude that the presence of $d$-states in valence is detrimental for the conservation of the graphene minicone.

\section{Conclusions}
In conclusion, this work confirms the possibility to induce, modify and dope the minicone state in graphene growth on Ni(111) by different kind of adatoms, revealing the microscopic mechanisms that determine the hybridization between Ni-$d$ states, graphene $p_z$ Dirac orbitals and RE valence electrons.
Our predictions for Eu adsorption on graphene are in line with what was recently experimentally observed for Eu adatoms on graphene/Co(0001)\cite{Jugovac2023eu} system, the main difference consisting on reduced doping in graphene due to the Ni(111) substrate.
Our extensive first-principles calculations, reveal the successfully doping of the spin-polarized minicone is realized when the adatoms do not present $d$-states in valence, while the presence of them  leads to the formation of a single-spin electron-like state crossing the Fermi level around the K-point of the graphene BZ which flattens at $\sim-0.25$~eV along the K-M direction.

Our work proposes a feasible way to engineer the minicone band present in graphene on Ni(111) substrate, which could serve as a material platform for spintronic applications, transport experiments and Kondo physics.\cite{graphene-Kondo, graphene-spintronics}.


\section*{Acknowledgements}

C. T. acknowledges financial support under the National Recovery and Resilience Plan (NRRP), Mission 4, Component 2, Investment 1.1, 
funded by the European Union – NextGenerationEU– Project Title “DARk-mattEr DEVIces for Low energy detection - DAREDEVIL” – CUP D53D23002960001 - Grant Assignment Decree No. 104  adopted on 02-02-2022 by the Italian Ministry of Ministry of University and Research (MUR).

C. T. and F. B. acknowledge financial support under the National Recovery and Resilience Plan (NRRP), Mission 4, Component 2, Investment 1.1, funded by the European Union – NextGenerationEU– Project Title "Symmetry-broken HEterostructurEs for Photovoltaic applications - SHEEP" – CUP B53D23028580001 - Grant Assignment Decree No. 1409  adopted on 14-09-2022 by the Italian Ministry of Ministry of University and Research (MUR).

Research at SPIN-CNR has been funded by the
European Union - NextGenerationEU under the Italian Ministry of University and Research (MUR) National Innovation Ecosystem grant ECS00000041 - VITALITY, C. T. acknowledges Università degli Studi
di Perugia and MUR for support within the project
Vitality.

G.P. acknowledges the European Union-NextGenerationEU under the Italian Ministry of University and Research (MUR) National Innovation Ecosystem Grant No. ECS00000041 VITALITY-CUP E13C22001060006 for funding the project.

C. T and G.P. acknowledge support from CINECA Supercomputing Center through the ISCRA project and Laboratori Nazionali del Gran Sasso for computational resources. 

\bibliography{graphene}{}


\end{document}